\documentclass[runningheads]{llncs}
\usepackage{fdsymbol}

\usepackage{graphicx}

\begin{document}

\title{A Survey on Game Theory Optimal Poker}
%
%
\author{Prathamesh Sonawane \and
Arav Chheda}
\authorrunning{P. Sonawane, A. Chheda}
%
\institute{University of Illinois Urbana-Champaign, USA\\
\email{pks10@illinois.edu, aravmc4@illinois.edu}\\
\url{}
}
\maketitle
\begin{abstract}
Poker is in the family of imperfect information games unlike other games such as chess, connect four, etc which are perfect information game instead. While many perfect information games have been solved, no non-trivial imperfect information game has been solved to date. This makes poker a great test bed for Artificial Intelligence research. In this paper we firstly compare Game theory optimal poker to Exploitative poker. Secondly, we discuss the intricacies of abstraction techniques, betting models, and specific strategies employed by successful poker bots like Tartanian\cite{no-limit-poker-sandholm} and Pluribus\cite{superhuman-ai}. Thirdly, we also explore 2-player vs multi-player games and the limitations that come when playing with more players. Finally, this paper discusses the role of machine learning and theoretical approaches in developing winning strategies and suggests future directions for this rapidly evolving field.

\end{abstract}
\section{Introduction}

Game Theory (GT) is an area of applied mathematics focused on examining the behavior of rational entities as they vie for a finite set of shared network resources. It aims to understand how these entities can cooperate or compete to achieve a balanced distribution of system resources that meets their individual needs. GT delves into the dynamics between independent and self-interested entities. Poker is a classic example of an imperfect information game because players don't know each other's cards. Recently, significant research has been directed towards the development of Game Theory Optimal (GTO) strategies in poker, alongside the creation of AI-driven agents. These agents also incorporate GTO principles to engage in play with human opponents in the game of Poker.
\newline\newline
In this study, we'll specifically be investigating the creation of agents and strategies that maximize winning potential. Given the exponential growth of the poker game-tree, we'll cover abstraction methods and discretized betting models that reduce the game complexity but still retain enough similarity to allow the strategies to be effective in real games. Additionally, we will discuss exploitative strategies as they allow for more profit in real world games while also requiring less abstraction to be effective. Furthermore, the paper will examine how development approaches need to be changed for games with multiple player compared to heads up play.

\section{Background}
In this paper, we will mainly cover Heads Up Limit Texas Hold'em Poker. In this version of poker, there are only two players. Each player receives two cards (known as the hole cards) which are known only to them. They then bet over 4 rounds. The first is the preflop, which takes place before any cards are on the board. The next round is the flop, which has 3 cards on the board. Next is the turn, which has an additional $4^{th}$ card on the board. Finally is the river which adds the $5^{th}$ and final card to the board. Each round, the player can either fold, call, or raise. The hand ends when either all but one player fold, or all 4 rounds are completed in which case the player with the best hand will take the money in the center. 
\newline\newline
One important variation is no-limit vs limit poker. In limit variations, the betting amounts are pre-determined and structured. For example, in a \$2/\$4 Limit game, the bets and raises on the first two rounds (pre-flop and flop) must be in increments of \$2, and the bets and raises on the last two rounds (turn and river) must be in increments of \$4. In addition, there is a limit to the number of raises allowed in a round (usually only 4). These changes help to reduce the complexity of the game compared to no-limit games where betting/raising is unrestricted. In limited versions of the game, there are 3.6 x $10^{17}$ possible variations in the game tree.

\section{Main Survey sections}
\subsection{Exploitative vs GTO}

GTO (Game Theory Optimal) poker is essentially “perfect poker”. This strategy is unexploitable regardless of the opponent's strategy. If all players at the table played GTO poker, a Nash Equilibrium would occur and everyone would generate 0 profit. It is important to note, however, that GTO poker is currently impossible to play because there are too many variables at play and calculating all of the probabilities and values without abstracting certain parts of the game is computationally impossible. Many solvers come close to approximating the values using clever abstraction techniques and reductions, but they have their limitations. While GTO poker guarantees at least a neutral EV, it has one major flaw: real-life opponents (especially human ones) are not able to play GTO poker. Opponents will often deviate from GTO poker unintentionally due to mistakes in their strategy--something that we can exploit to further increase our EV (even if we have to deviate from a GTO gameplan). 

This is exactly where Exploitative poker comes in. Exploitative poker deviates from GTO poker play in order to take advantage of the mistakes your opponent is making. One such example of an exploitative bot is Loki\cite{Loki}. Loki is a poker program capable of observing its opponents, constructing opponent models, and dynamically adapting its play to best exploit patterns in the opponents’ play. The authors behind Loki developed it with the following key components in mind.
\begin{itemize}
    \item \textbf{Assessing hand strength}: How strong your hand is in relation to the other hands.
    \item \textbf{Assessing hand potential}: The probability of a hand improving (or being overtaken) 
    \item \textbf{Betting strategy}
    \item \textbf{Bluffing strategy}
    \item \textbf{Unpredictability}
    \item \textbf{Opponent modeling}
\end{itemize}
We will further discuss how the Loki bot works, but a key point is that the Loki bot has a default strategy for aspects like betting and bluffing, however, this strategy slowly changes as Loki develops models for its opponents. Starting of with a GTO game plan and then tweaking it as more opponent data is gathered is a common trend in poker bots


\subsection{Abstraction Techniques}
\begin{figure} [!htp]
    \centering
    \includegraphics[width=1.0\textwidth]{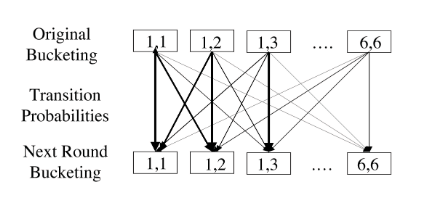}
    \caption{Transition probabilities visualized \cite{approximating-poker}.}
    \label{fig:transition}
\end{figure}

Poker is an extremely complex game. Due to the immense scale and complexity, developing purely theoretical solutions for the full game is not possible with current technology and approaches. Instead, researchers attempt to create a simplified version of the game that still retains enough of the original complexity for the strategy to successfully transfer from the abstracted game to the real game.
We will look at the following paper:  "Approximating Game-Theoretic Optimal Strategies for Full-scale Poker" by D. Billings \cite{approximating-poker}. This paper details several methods for reducing the complexity of heads-up Texas Hold'em poker to develop a competitive poker bot. An initial approach the paper highlights is card reduction, which involves decreasing the number of cards in play—whether in the deck, the player's hands, or on the board. This reduction helps create smaller and simpler versions of the game for testing abstraction techniques and comparing them with near-optimal solutions. However, these simplified games can deviate significantly from actual Texas Hold'em, limiting their usefulness for approximating full-scale play.
\newline\newline    
Another strategy is modifying the betting round structure. This includes limiting players to a maximum of three actions per round, which reflects typical gameplay where many hands don't exceed this number of bets (regardless of whether the game is limit or no-limit). Another option is the elimination of later betting rounds, which substantially shrinks the game tree. While this simplification makes the game more manageable computationally, it also leads to strategies that may not perform as well when applied to the full game. It is important to note, however, that the impact of this simplification is less severe for early rounds, allowing for the development of effective early-round strategies based on the reduced model.
Additionally, to address the truncation of betting rounds, the paper suggests using expected value calculations at the leaf nodes of the game tree. This method averages the potential outcomes of the truncated rounds, rather than prematurely declaring a winner based on the current strongest hand. For example, in a 3-round model, all possible upcoming 4th round cards are considered to calculate the expected value. By considering all possible outcomes for the next round, the model maintains a degree of complexity that accounts for the rounds that have been removed. A flaw in this approach, however, is that it can only consider up to one bet per player in the missing rounds. Many times these later rounds can have multiple bets per player, meaning that the bot’s strategy will miss some nuances of the actual game's strategy.
\newline\newline
Another way to reduce the complexity of modeling poker is to simplify how prior rounds are modeled. If we took a fully GTO approach, we would have to consider the game tree nodes as dependent on the actions taken during the prior round. However, this paper also details several ways to simplify this while retaining much of the key complexity. The first is to disregard all prior probabilities in favor of a uniform distribution over all possible hands assumed for each player. Different post-flop solutions are then computed based on initial pot sizes that correspond to various preflop betting sequences without needing to consider who specifically raised and called the bet.
The paper then discusses a more advanced approach where they estimate the conditional probabilities using the solution to a preflop model. For example, in the PsOpti2 program discussed in the paper, a preflop model was developed and solved, leading to a pseudo-optimal preflop strategy. The resulting distributions then were used as input parameters for the post-flop models. By assuming that all players would be following this pseudo-optimal preflop strategy, the post-flop model was able to generate an effective approximation of the full game's perfect recall knowledge at much less computational cost.
\newline\newline
The next, and most important, method of abstraction discussed by the authors is bucketing. The concept of "bucketing" in poker strategy computation is an abstraction method that simplifies the game's complexity by categorizing hands into groups, or buckets, based on their strategic similarity. This process involves creating equivalence classes where each class consists of hands that can be played in a similar fashion without significantly impacting EV. The bucketing considers both the current hand strength and the hand's potential to improve (For example the hands that are weak but have a good chance of completing a flush or straight). The authors finally decided to allocate all but one bucket based on hand strength with an additional bucket for high-potential hands suitable for semi-bluffing strategies. Each bucket contained a different number of potential hands, with weaker buckets containing a larger range of hands than stronger buckets. Both the player’s hand and the opponent’s hand are modeled using bucketing, with up to 7 buckets per player. 
In poker, the strength of hands can wildly change from round to round. This is where transition probabilities are used. Transition probabilities represent the likelihood of moving from one bucket to another as new community cards are revealed. We consider every card in the deck and look at how it affects each bucket. We consider how these new cards will affect both our and our opponent’s buckets meaning an n x n transition network is created (Figure \ref{fig:transition}). The abstract game is then solved using linear programming, where the objective function is to maximize EV, which generates a table of mixed strategies for each possible game scenario. A poker-playing program uses this table, applying measures of hand strength and potential to select an action from the mixed strategy that corresponds to the hand it's dealt. Although the LP computations are resource-intensive, typically requiring a day and substantial memory, they are feasible and produce a usable strategy for the program (provided the number of buckets is sufficiently small), streamlining the decision-making process to a level that is computationally manageable while still being effective.
\subsection{Betting models}
\subsubsection{Discretized betting model\cite{no-limit-poker-sandholm}:}
This is the betting model for the agent Tartanian which placed second out of ten entries at the AAAI-07 Computer Poker Competition. \\

In limit Texas Hold’em Poker, the players only ever have at most three possible actions available to them (fold, call, or raise), but in no-limit Texas Hold’em Poker players can choose to do any of the infinite set of plays at their disposal. For example, an agent can fold, call, or raise to any (integral) amount between the minimum amount and his budget, and if the bets are not limited to integral amounts then the branching factor would actually be infinite.\\

Since the betting space is very huge they bucket the betting amounts and use a discretized betting model. This makes sense because in practice some of the betting amounts are made very often compared to others. For ex- bets equal to half of the size of the current pot, bets equal to the size of the current pot, and all-in bets. We don't include small bets made with respect to the pot because that is typically disadvantageous. When an opponent encounters a small bet, they receive highly favorable pot odds to call, allowing them to continue in the hand with a wide range of possible hands. Additionally, for a player holding a strong hand, such minimal bets contribute insignificantly to building the pot's value. Of course, this excludes the case where the player only has a small amount to bet and is essentially going all-in.\\

Essentially this is how the betting model works:
\begin{itemize}
    \item When the agent is initiating a round, the actions available are check, bet half the pot, bet the pot or go all-in.
    \item When the agent is responding to bets placed in a round, the actions available are fold, call, bet the pot, or go all-in.
    \item If at any point a bet of a certain size would commit more than half of a player’s stack, that particular bet is removed from the betting model.
    \item At most three bets (of any size) are allowed within any betting round.
\end{itemize}
Now to map this betting model to real-life scenarios the authors had to decide how they would handle situations where the opponents bet if outside the predecided set of discretized betting amounts (something like $\frac{3}{4}^ths$ of the pot size.). In this case, they first tried to perform rounding but soon discovered that this can be exploited as the opponent can then predict whether the agent will bet or not based on their betting amount. Hence they developed a randomized weighted metric based on relative distance from discretized amounts to determine what the agent's betting amount must be. Given the opponent put in c chips in the pot and $d_1$ and $d_2$ are the discretized betting amounts corresponding to c s.t. $d_1 < c < d_2$, their agent would then choose the amount corresponding to the minimum of $\frac{c}{d_1}$ and $\frac{d_2}{c}$

\subsubsection{Betting model in Loki\cite{Loki}:}
We'll now discuss the hand assessment techniques used in the bot Loki and subsequently the betting model used. There are 4 major aspects to hand assessment that the authors touch upon.
\begin{itemize}
    \item \textbf{Preflop Evaluation: } In poker, there are 1,326 possible hand combinations from a 52-card deck, but effectively only 169 unique hand types. The researchers simulated 1 million games to calculate an income rate (profit expectation), which helped determine their preflop value.

    \item \textbf{Hand strength: } The probability of having the best hand in poker during the flop, turn, \& river stages can be estimated using enumeration methods. This involves counting hands that are better, equal, or worse than ours, without considering opponent strategies.

    \item \textbf{Hand Potential: } In poker, a hand with initially low strength, like $5\heartsuit2\heartsuit$, can significantly improve with the right community cards. To evaluate this, we consider the 990 combinations of the next two cards against 1,081 opponent hands, counting situations where we lead, tie, or trail. This approach, while informative, requires adjustments as not all hands are equally likely.

    \item \textbf{Betting Strategy: } Pot odds are your winning chances compared to the expected return from the pot. For example, if you assess your chance of winning to be 25\%, you would call a \$4 bet to win a \$16 pot (4/(16+4) = 0.20) because the pot odds are in your favor (0.25 $\ge$ 0.20). The formula that they use is $EHS = HSn + (1 - HSn) * Ppot$
\end{itemize}

\subsection{Exploitation Strategies}
\subsubsection{CFR+ \cite{limit-holdem-solved}}

In this paper, the researchers have managed to create an algorithm that essentially "weakly solves" Heads-Up Limit Texas Hold'em. They created a strategy that will always at least tie, if not win, in the long run against any opponent. They also prove the long-held notion that the dealer in poker has an advantage. 

\begin{figure} [!htp]
    \centering
    \includegraphics[width=1.0\textwidth]{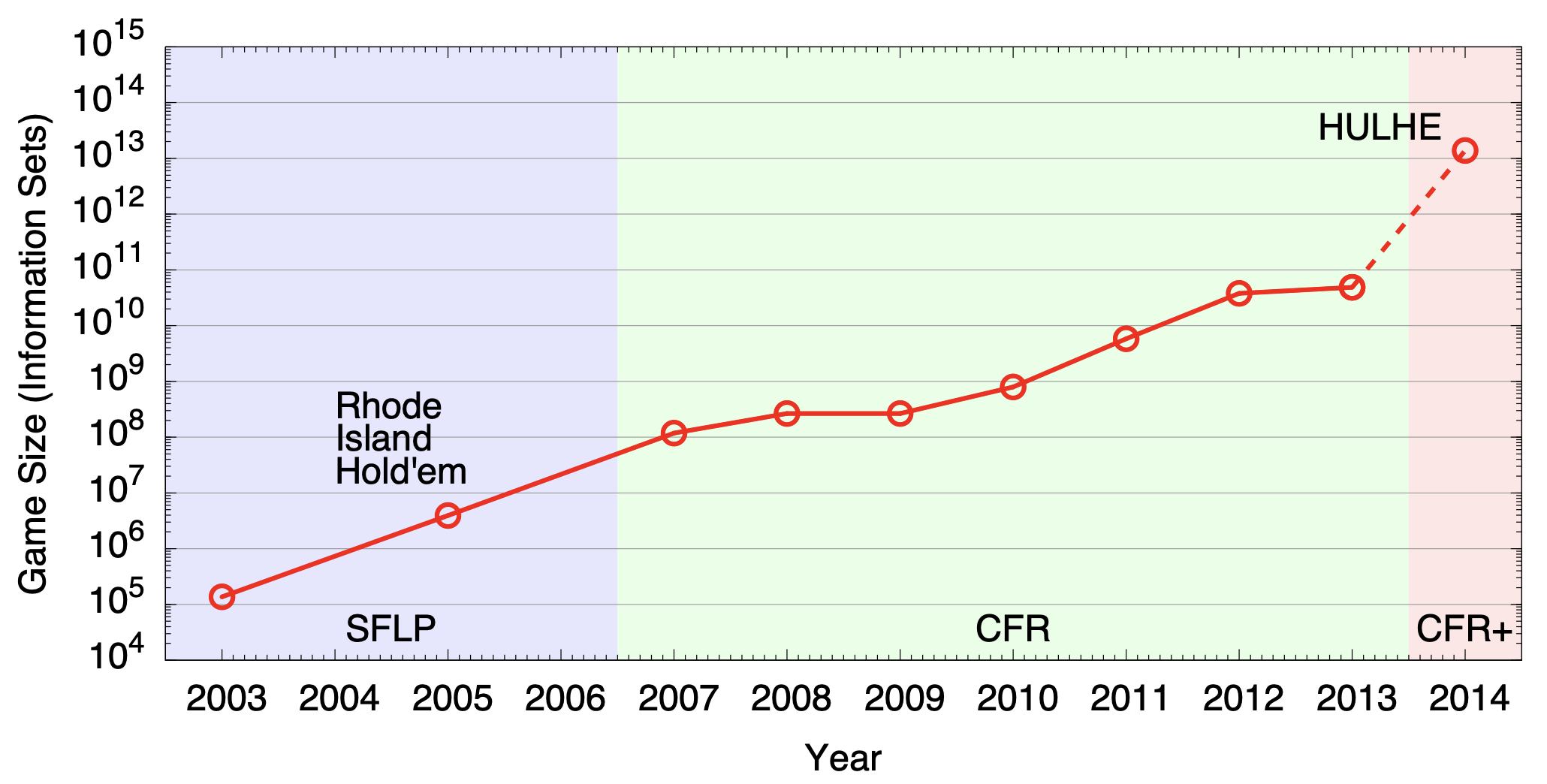}
    \caption{Increasing sizes of imperfect-information games solved over time measured in unique information sets (i.e., after symmetries are removed). With the algorithm discussed (CFR+) being able to solve games of size upto $10^{13}$\cite{limit-holdem-solved}}
    \label{fig:CFR}
\end{figure}

This algorithm is called CFR+ (Counterfactual Regret Minimization Plus). It can handle much larger games than previous algorithms \ref{fig:CFR} and works by playing the game against itself, learning over time which strategies work best. It minimizes "regret" from not having played the best strategy in hindsight. By minimizing regret, CFR aims to converge to a strategy that can be minimally exploited by an opponent. 

For example, if you are playing a game where you have to choose between two actions: A or B. If you choose A, you win 1\$, and if you choose B, you win 2\$. Let's say you didn't know the outcomes beforehand, and you chose A. Later, you find out that B would have given you 2\$. The regret for not choosing B when you chose A is the difference in utility, i.e. 1\$. So, your regret for that decision point is 1\$.

In CFR, this concept is expanded to include a cumulative total of regret for each action over many iterations of playing against itself, guiding the algorithm toward strategies that would have won more in the past. The CFR+ algorithm, as used in the paper, improves upon this by ensuring that regrets are always non-negative and can rapidly update strategies by using regret values to make better decisions in future iterations. In CFR+, exploitability, which is a measure of how much the strategy can be exploited by an opponent, was reduced to 0.986 milli-big-blinds per game (HULHE), which is effectively negligible.

\subsection{2 Player Vs N Player}
\begin{figure} [!htp]
    \centering
    \includegraphics[width=1.0\textwidth]{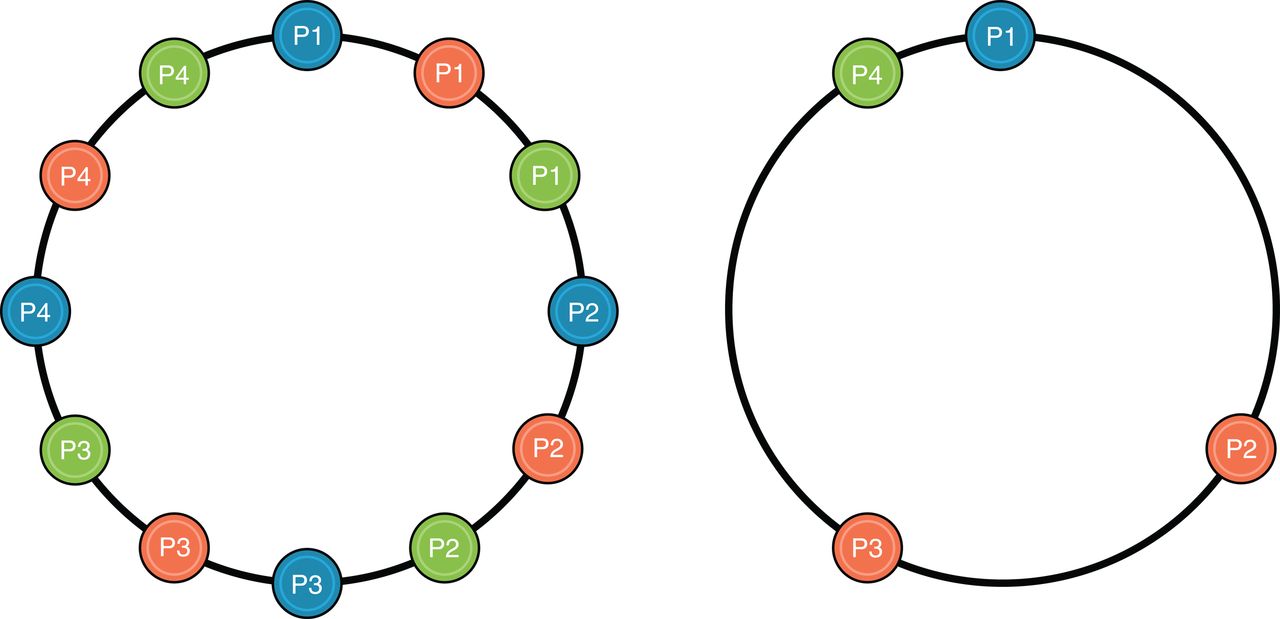}
    \caption{Example of lemonade stand game \cite{superhuman-ai}. In this game, participants choose a location on a circular space, aiming to maximize distance from others. The game has numerous Nash equilibria, each representing uniform player distribution around the circle. However, when players individually select different equilibria without coordination, the resulting strategy typically does not form a Nash equilibrium. This concept is depicted through illustrations: the left shows three equilibria using distinct colors, while the right demonstrates the non-equilibrium outcome when players independently choose different equilibria.}
    \label{fig:lemonade}
\end{figure}

Current poker (as well as most other games, such as chess and go) bots almost exclusively deal with solving two-player games. However, there has been research done on expanding this to games with more than two players. One example of this is discussed in "Superhuman AI for Multiplayer Poker" by Noam Brown and Tuomas Sandholm. In this paper, they discuss their work on “Pluribus” a bot that aims to beat its opponents in 6-player Texas Hold’em games. We will analyze this paper to see what compromises were made when it came to solving 6-player games rather than 2-player games.
\newline\newline
As mentioned earlier, in two-player games, finding a Nash Equilibrium allows for poker play that is unexploitable but is difficult to compute. When it comes to six-player poker, the complexity only further increases and a Nash equilibrium is even more computationally challenging to find. Currently, there is no known way to compute the Nash equilibria of a 2 player non-zero-sum game in polynomial time (Proving this is beyond the scope of this survey, but proof can be found here: X. Chen, X. Deng, S.-H. Teng, Settling the complexity of computing two-player Nash equilibria. J. Assoc. Comput. Mach. 56, 14 (2009)). We can also conclude that solving 3+ player zero-sum games must be just as hard, as we can always convert a 2-player non-zero sum game into a 3+ player zero-sum game by adding dummy players that receive the necessary payoffs to make the game zero-sum. Furthermore, even if a Nash equilibrium were found, it might not be the most effective strategy since the joint strategy of all players choosing their individual Nash equilibria does not necessarily result in a collective Nash equilibrium. This is due to the multitude of equilibria that can exist in multiplayer settings. An easy example of this is The lemonade stand game competition (Figure \ref{fig:lemonade}), devised by M. A. Zinkevich, M. Bowling, M. Wunder. This game has players place themselves in a ring with the goal of being as far apart from one another as possible. It is obvious that the Nash equilibrium is for all players to be uniformly spaced apart, and since there are an infinite number of ways to do this, there are an infinite number of Nash equilibria. However, this is unlikely to happen when all players compute the Nash equilibria independently. Only in 2 player zero-sum games does computing a Nash equilibria independently result in a Nash equilibria for both players.
\newline\newline
As a result of this, Pluribus was not developed to adhere strictly to GTO play or to find a Nash equilibrium using a theoretical approach. Instead, it uses a machine learning-based approach that, while not GTO, proved to be highly effective in practice, beating many top players and was profitable against many other bots and top human players. In summary, due to the extreme difficulty in computing the relevant Nash equilibria in 2+ player games, poker bots such as Pluribus that are aimed at solving multiway games must stray away from theoretic, symbolic models in favor of exploitative, neural models instead. 



\subsection{Limitations/Future work}


Although calculating the most optimal move for GTO poker is complex and an impossible task right now, we can look forward to a time when calculating those values will be possible. Regardless, throughout the paper we discussed popular design decisions made by authors to construct a poker bot, like discretized betting models, and abstraction techniques, all in an attempt to reduce the size of the decision tree. A promising area of work they mentioned is to automate the determination of these parameters (ex- determining the number of buckets and bucket size in the discretized betting model. We could use the amount lost by eliminating strategies to guide these decisions).

As for Exploitative poker, research in solving imperfect information games using AI is moving at a fast pace right now. Recent State-of-the-art bots like Rebel\cite{rebel-poker-bot}, and AlphaHoldem \cite{alphaholdem-poker-bot} use reinforcement learning and search to beat previous mill-big-blinds per game records in 2-player games.

While we understand that when developing poker bots, the goal is often not to actually win but to create new strategies and techniques, when it comes to purely winning money in the widest range of competition possible we recommend that individuals go for a machine learning approach over a more theoretical one. Firstly, ML approaches tend to be easier to compute meaning they require fewer game abstractions when coming up with a solution, meaning that strategies will transfer over well to real games. Additionally, ML approaches are effective in games with more than two players, something theoretical approaches struggle with. Machine learning-based approaches are also able to learn from their opponent's behavior and adjust, allowing for exploitative play, which generates more profit than a GTO game style against human opponents. If a purely ML approach is not preferred, we recommend at least a hybrid approach that incorporates some level of opponent analysis and adjustment (something that many top theoretical-based poker bots do).

\newpage
\section{Appendix}
\begin{figure} []
    \centering
    \includegraphics[width=1.0\textwidth]{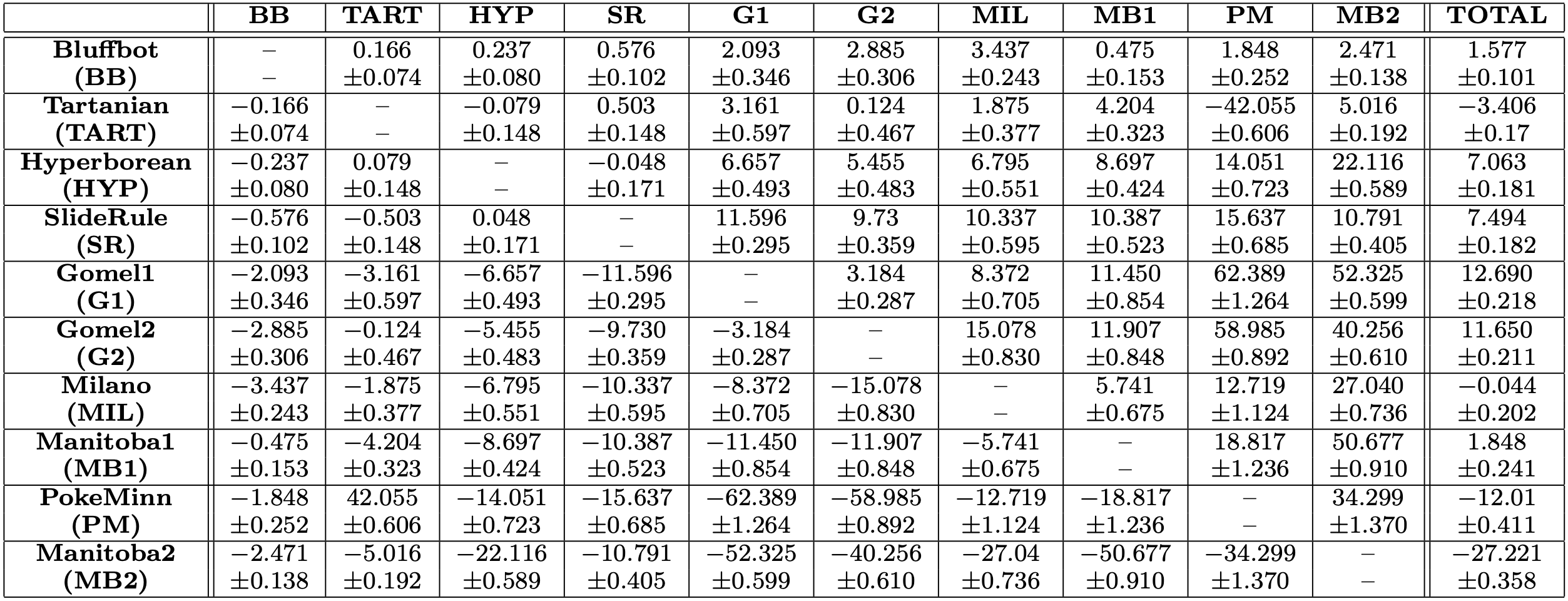}
    \caption{Results from the 2007 AAAI Computer Poker Competition. The players are listed in the order in which they placed in that competition. Each cell contains the average number of chips won by the player in the corresponding row against the player in the corresponding column, as well as the standard deviation. The numbers in the table reflect 20 pairwise matches each; in the AAAI competition a further 280 matches were conducted between each pair of the three top-ranked entries in order to get statistical significance, and Tartanian finished second.\cite{no-limit-poker-sandholm}}
    \label{fig:tart}
\end{figure}

\end{document}